# Thermionic Cooling of the Target Plasma to a Sub-eV Temperature

M. D. Campanell* and G. R. Johnson
*Lawrence Livermore National Laboratory, P.O. Box 808, Livermore, CA 94551, USA*
*michaelcampanell@gmail.com

Contemporary models of bounded plasmas assume that the target plasma electron temperature far exceeds the temperature of the cold electrons emitted from the target, $T_{emit}$. We show that when the sheath facing a collisional plasma becomes inverted, the target plasma electron temperature has to equal $T_{emit}$ even if the upstream plasma is hotter by orders of magnitude. This extreme cooling effect can alter the plasma properties and the heat transmission to thermionically emitting surfaces in many applications. It also opens a possibility of using thermionic divertor plates to induce detachment in tokamaks.

Thermionic electrons are emitted from plasma-facing surfaces in many systems. Important examples include tokamak divertors [1], hypersonic vehicles [2], dusty plasmas [3], emissive probes [4,5], the Large Plasma Device [6], thermionic discharges [7], and arcs [8]. Emission alters the sheath potential and energy balance, thereby affecting the interior plasma properties [9]. Conventional treatments of emission [4,10] maintain that $T_{et}$, the plasma electron temperature near the target (surface), can far exceed the sub-eV temperature of thermionic electrons $T_{emit}$. This seems reasonable since laboratory plasma sources tend to have electron temperatures ranging from a few eV [6] to hundreds of eV [11] or more. But in this Letter, we show that strong emission causes a sharp reduction of $T_{et}$ to $T_{emit}$.

Contemporary models are based on the assumption that the target sheath is classical under weak emission and space-charge limited (SCL) [4,10] under strong emission, see Fig. 1(a). In either case, most plasma electrons approaching the target are reflected. The sheath accelerates the thermionic beam, making it have a low density relative to the other electrons at the sheath edge, as shown in the electron velocity distribution function (EVDF) in Fig. 1(b). The electron population near the target is therefore dominated by confined plasma electrons. That is why the target plasma was allowed to have a high $T_{et}$ that was independent of $T_{emit}$.

Recent works [12,13] find that SCL sheaths are destroyed by ion trapping in the virtual cathode (VC) and the equilibrium strongly emitting sheath must be inverse, see Fig. 1(a). Inverse sheath formation was recently shown to trigger temperature saturation in Hall thrusters [14], mode transitions in thermionic discharges [15], and ion-ion plasma formation in negative ion sources [13]. Inverse sheath theory also explains why emissive probes often float above the plasma potential [5].

The influence of emission on the plasma temperature and heat transport needs to be reconsidered to account for the inverse sheath effect. This Letter will show that the only way to patch an inverse sheath to a collisional plasma is for the target plasma electron temperature $T_{et}$ (temperature in the quasineutral region near the sheath edge) to be $T_{emit}$. We start from the usual assumption that thermoelectrons are emitted with a half-Maxwellian distribution at temperature $T_{emit}$ [10]. Because an inverse sheath is a monotonic potential barrier to the thermoelectrons, those entering the plasma from the sheath edge will also be half-Maxwellian with temperature $T_{emit}$. Because the EVDF in a collisional plasma must be roughly a full Maxwellian [9], we conclude that the target-directed plasma electrons near the inverse sheath edge will have a thermal spread corresponding to $\sim T_{emit}$ as well.

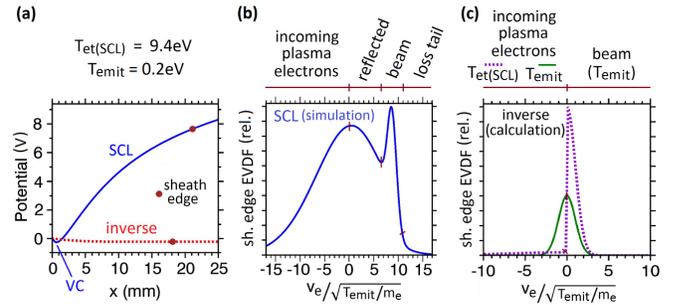

FIG. 1. (a) Representative sheath potentials from simulations (detailed later). The SCL confines hot plasma electrons. The EVDF at the SCL sheath edge (b) shows that most electron velocities far exceed the thermal velocity of emission. Inverse regimes differ because the plasma electrons are unconfined and beam electrons enter with low velocities. Panel (c) shows calculated EVDF's at an inverse sheath edge assuming incoming plasma electrons have (i) hot temperature $T_{et(SCL)}$ *and* (ii) temperature $T_{emit}$. As we will discuss, assumption (i) leads to a paradox while (ii) leads to a full Maxwellian EVDF that patches smoothly to a collisional plasma.

To further clarify this result, let us suppose instead that the target-directed electrons at the inverse sheath edge had a hot temperature. Assuming the surface draws zero current and ions are confined, the oppositely directed hot and cold electrons must carry equal fluxes, requiring their partial densities to satisfy $n_{hot}T_{hot}^{1/2} = n_{cold}T_{emit}^{1/2}$. Thus $n_{cold} \gg n_{hot}$ and the inverse sheath edge EVDF would have the lopsided form in the dotted curve of Fig. 1(c). This is paradoxical as the hot electrons should have been cooled by collisional energy exchange with the denser cloud of cold electrons. The paradox vanishes if the target-directed electrons also have temperature $T_{emit}$, which will be the case in the collisional limit.

In the remainder of this Letter, we will show the formation of inverted cold target plasmas in simulations, set corresponding boundary conditions for fluid models, and suggest a potential application to mitigate plasma-surface interactions in tokamaks. In previous simulations of emission effects by many authors [1,10,16,17] including our Ref. [12], the temperature of plasma electrons bombarding the target was constrained to a high value in the setup. No simulation studies have analyzed how emission feedback cools plasma electrons. Experimental studies [5,18,19] have focused on measuring sheath properties, so the cooling of the plasma interior is not as well understood. Our new code is designed to study the cooling effect at a fundamental level. It solves the continuum kinetic [20] Boltzmann transport equation (1) on a discrete 1D-1V grid. The distribution functions of the electron and singly charged deuterium ion species $f_{e,i}(x,v_{e,i})$ are advanced until equilibrium is reached.

$$\frac{\partial f_s}{\partial t} = -v_s \frac{\partial f_s}{\partial x} + \frac{q_s}{m_s}\frac{\partial \varphi}{\partial x}\frac{\partial f_s}{\partial v_s} + S_{coll(s)} + S_{charge(s)} \quad (1)$$

The simulated domain is $L = 1m$ long. The left boundary is the surface and the right boundary is a symmetry plane. To simulate electron thermalization and heating, a charge-conserving BGK [21] operator perturbs $f_e$ towards a Maxwellian at a rate set by $e_{mfp}$, an effective mean free path of electrons with thermal velocity $(T_e/m_e)^{1/2}$.

$$S_{coll(e)} = \frac{\sqrt{T_e/m_e}}{e_{mfp}}\left[ n_e \sqrt{\frac{m_e}{2\pi T_e}} \exp\left(\frac{-m_e v_e^2}{2T_e}\right) - f_e \right] \quad (2)$$

As diagrammed in Fig. 2, there is a 0.15m "heating region" upstream. Upstream heating is used to resemble a typical system where hot plasma electrons are produced at a distance from the surface, such as the tokamak scrape-off layer (SOL) [9]. In the heating region, operator (2) uses a set temperature $T_{eup} = 20eV$ and $e_{mfp} = 0.03m$. This $e_{mfp}$ value causes enough heating to keep the peak electron temperature in the heating region near 20eV in all simulations presented.

Outside the heating region, (2) uses the local effective temperature $T_e(x)$ calculated each time step from the mean energy. This treatment captures the energy-conserving nature of e-e self-collisions and gives the flexibility to adjust the collisionality by varying $e_{mfp}$. The regimes of interest here have a Knudsen number [22] $e_{mfp}/L \ll 1$. To focus on the cooling of plasma electrons by emission, the complex exchange of electron energy with ions and neutrals is not simulated, nor would it alter the main results to be shown. BGK operators for interspecies collisions [21] can be employed in future analyses. With the present code, one can study how the $T_e$ profile in a plasma depends on the emission intensity, the collisionality, and the sheath type.

The volumetric charge source $S_{charge}$ produces ion-electron pairs to offset ion losses to the surface, keeping the mean ion density $\langle N \rangle$ fixed. Sourcing is uniform in space for simplicity. Source ions have temperature $T_i = 0.2eV$. Source electrons are produced at the local temperature $T_e(x)$ in order to not alter $T_e$. Ions suffer charge-exchange (CX) collisions with mean free path 0.2m. The CX collisions replace ions (which gain energy accelerating in the potential) with ions at $T_i = 0.2eV$. Thermoelectrons are emitted with a half-Maxwellian EVDF at $T_{emit} = 0.2eV = 2321K$. The surface is floating and the emission coefficient $\gamma$ (ratio of emitted flux to the influx of plasma electrons [10]) is controlled.

Fig. 2 shows how electron emission affects the $T_e$ profile in representative runs. Starting from the classical sheath $\gamma = 0$ case, one observes that increasing $\gamma$ reduces the target electron temperature $T_{et}$ until the sheath potential becomes SCL. In this example, $T_{et}$ at the SCL limit is 47 times larger than $T_{emit}$. Increased emission beyond the SCL limit is suppressed by a growing VC, so the $\gamma = 2$ and $\gamma = 1$ SCL profiles in essence overlap.

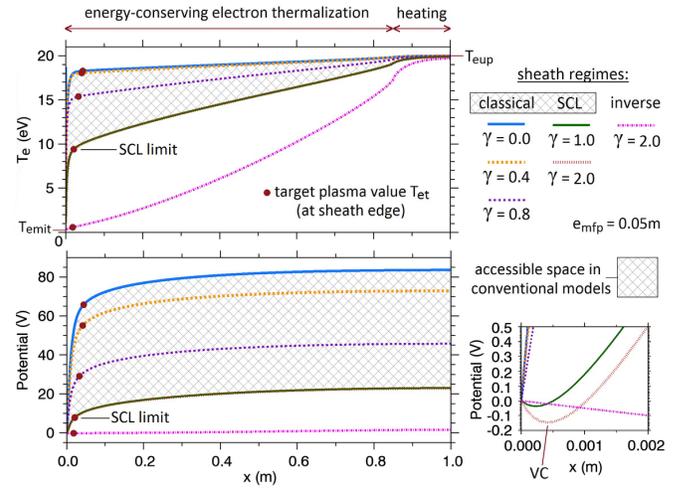

FIG. 2. Dependence of the $T_e$ and potential profiles on emission intensity in the classical, SCL and inverse sheath regimes.

SCL states are simulated here to illustrate conventional predictions but are unlikely to exist in practice. Recent studies show that CX collisions make cold ions accumulate in the VC, causing the sheath to invert [12]. This is why collisionless simulations of emitting divertor sheaths by Komm et al. [1] produced stable SCL's while simulations of H- emission with CX collisions by Zhang et al. [13] yielded an inverse regime. SCL equilibria here are sustained by excluding the CX collisions and $S_{charge}$ in the VC potential well. The inverse equilibria have similar control parameters but no exclusions. Also, because inverse sheaths are far thinner at a given plasma density, $\langle N \rangle$ is set to $5 \times 10^{10}$ m$^{-3}$ for inverse cases versus $10^{14}$ m$^{-3}$ for classical and SCL. This way, all sheaths produced have similar widths and are thin versus L. Resolving the sheaths is computationally nontrivial even in 1D. The SCL cases use 10,000 spatial cells to ensure good resolution of the VC and 400 velocity cells for each species to well resolve the peaks in velocity space.

Fig. 2 contains an inverse sheath case with the same $\{T_{eup}, e_{mfp}\}$ as the other cases, but $T_e$ drops to a sub-eV level near the target. Major differences between the plasmas in the SCL and inverse sheath equilibria are evident in Fig. 3. The SCL $f_i$ shows a sharp acceleration of ions towards the target. The inverse $f_i$ shows minimal directional flow because most ions are confined. In the SCL $f_e$ plot, most of the electrons near the target, including the thermionic beam, have energies far exceeding 1eV. The inverse $f_e$ features no energetic electrons near the target.

The extreme cooling ability of inverse sheaths was not discovered in previous studies of inverse regimes [12,14] because hot electrons were produced throughout the simulated plasma domain. Creation of hot electrons within the cold electrons near the target formed a two-temperature EVDF [12] like the lopsided dotted curve in Fig. 1(c). A lopsided EVDF can trigger strong instabilities related to the "self-spikes" observed in the anode glow mode of weakly collisional thermionic discharges [7,15].

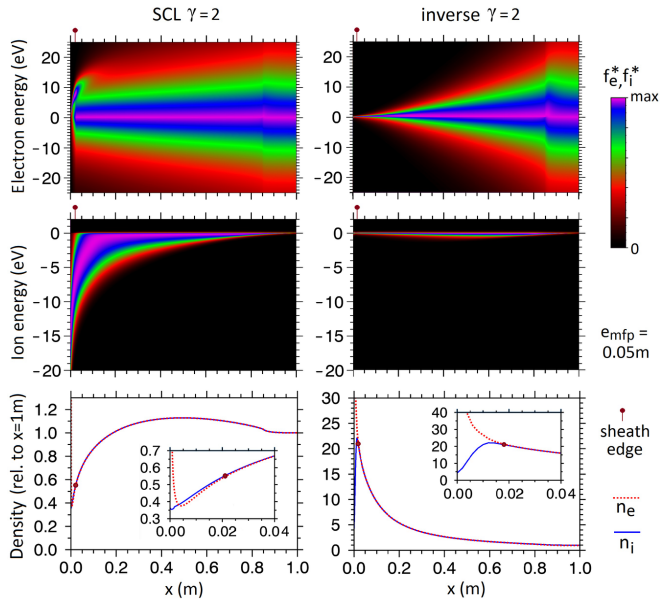

FIG. 3. Comparison of the SCL and inverse regimes. The color plots show the distribution functions normalized to the local density $f^*_s \equiv f_s(x,v_s)/n_s(x)$. Velocity is converted to energy coordinates (negative indicates motion towards the target). Charge density distributions are plotted at bottom. In all figures, the sheath edges are determined by the location where $n_e$ separates from $n_i$ by 1%.

This study shows that all electrons near an inverse sheath will be cold if the plasma's hot electrons are born at a distance well exceeding the mean free path of e-e collisions. The inverse $f_e$ in Fig. 3 is roughly Maxwellian at all x with a $T_e$ that drops in a smooth way from $T_{eup}$ upstream down to $\sim T_{emit}$ at the sheath edge. The quasineutral density increases a surprising 21 times from upstream to the sheath edge because the electron pressure $n_e T_e$ is almost conserved as $T_e$ drops. It is not exactly conserved due to the role of trapped ions in the force balance.

Next, we will compare how electron emission cools the plasma and alters heat flow in each sheath regime. In general, electron heat flow through a collisional plasma is related to the $T_e$ gradient [23]. Fluid models express the heat flow as $Q_{e,plasma} = -K dT_e/dx$, where K is a heat conduction coefficient that depends on plasma properties and collisions [9].

The $Q_{e,plasma}$ must (roughly) balance the electron heat flux carried through the sheath (neglecting heat sources or sinks such as energy exchange with heavy particles). As is well known, increasing $\gamma$ weakens a classical sheath potential and enhances transmission of hot plasma electrons to the surface [9,10]. To compensate, $Q_{e,plasma}$ must rise so a larger $T_e$ gradient needs to form, leading to a reduced $T_{et}$. This explains how increased emission cools the target plasma in the classical regimes of Fig. 2. The plasma is not cooled by losing energy to thermoelectrons (a common misconception), since the thermoelectrons accelerate into the plasma with energy from the sheath potential which may exceed $T_{et}$.

When the sheath is SCL, the target plasma cannot be further cooled by increased emission, as was shown in Fig. 2. Other ways to reduce $T_{et}$ in a classical or SCL regime are to add energy sinks (e.g. radiating impurities) or to operate at higher collisionality [9]. The influence of electron collisions on $T_e$ profiles is studied in Fig. 4. It is instructive to include effective temperatures of the electrons moving left and right ($T_{eL}$, $T_{eR}$) calculated from their mean energy. The result $T_{eL} > T_{eR}$ shows that heat is flowing toward the target. It is evident that collisions impede heat flow from the fact that raising the collisionality (reducing $e_{mfp}$) causes $T_{eL}$ and $T_{eR}$ to converge with $T_e$, bringing the EVDF closer to a true Maxwellian.

In the SCL regime, electron heat flow through the sheath is dominated by the loss of high energy electrons from the plasma. When $e_{mfp}$ is reduced, the heat conductivity K in $Q_{e,plasma} = -K dT_e/dx$ is reduced, so a higher $T_e$ gradient through the plasma is needed to conduct the lost heat. This explains the reduction of the target $T_e$ with reduced $e_{mfp}$ in the SCL cases of Fig. 4. A similar outcome was shown in the classical regime simulations by Tang and Guo [23]. Overall, higher collisionality can lower the target plasma temperature when the sheath is classical or SCL, but it is difficult to reach a sub-eV level in these regimes.

The inverse regimes in Fig. 4 show a unique, remarkable property that $T_{et}$ is always sub-eV, independent of $e_{mfp}$. The basic reason is that inverse sheaths flood the quasineutral plasma with unaccelerated thermoelectrons at their original temperature $T_{emit}$, establishing $T_{eRt} = T_{emit}$ exactly. Meanwhile, e-e collisions act to establish $T_{eL} \approx T_{eR}$ throughout the plasma. Combining these requirements, the target plasma EVDF becomes approximately a full Maxwellian with cold temperature $T_{emit}$ when $e_{mfp} \ll L$. Hot electrons born upstream can only bombard the target in the weakly collisional regime, $e_{mfp} > L$. In plasmas with weak e-e collisionality such as partially ionized low pressure plasmas [15,24], the EVDF can deviate far from Maxwellian such that an electron "temperature" is not well defined.

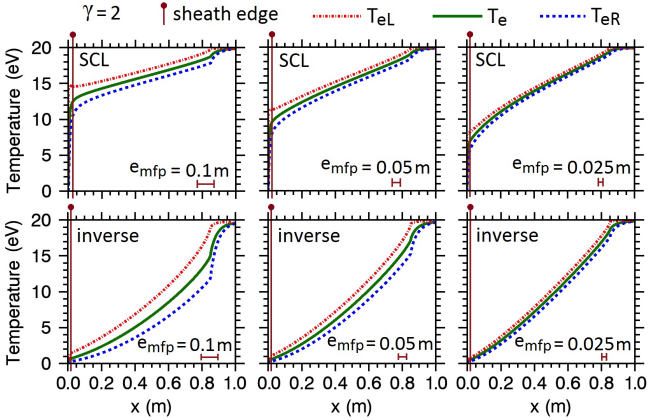

FIG. 4. Distributions of $T_e$ and one-way temperatures $\{T_{eL}, T_{eR}\}$ in plasmas bounded by SCL and inverse sheaths. The effect of electron collisions is shown by varying $e_{mfp}$ in the energy-conserving collision region $0 < x < 0.85$m. The data show that a very small $e_{mfp}$ is required to achieve a low target plasma electron temperature in the SCL regime, but not in the inverse regime.

A corollary of our analysis is that inverse sheaths can be patched to fluid models of collisional plasmas contacting strongly emitting surfaces by setting $T_{et}$ to $T_{emit}$ as a boundary condition. This differs from conventional boundary conditions where $T_{et}$ was allowed to be hot and SCL sheath energy transmission factors were employed [25]. Meanwhile, the conventional boundary condition on flow velocity was the ion sound speed from the Bohm criterion [26]. For the inverse regime, the boundary velocity can be assumed zero if ions are well-confined. In addition to $T_{et} = T_{emit}$, the ions (and neutrals) can also be assumed to have temperature $T_{emit}$ at the target in collisional inverse regimes, even if the upstream ions are hot. This is because high interspecies collisionality leads to equal temperature profiles among particle species [9]. Ions at temperature $T_{emit}$ will then be well-confined by the inverse sheath potential barrier $q_e\Phi_{inv} = T_{emit}\ln(\gamma)$ [12] if $\gamma$ is well above unity.

Scenarios with an inverse sheath and cold target plasma are feasible not only with thermionic emission but also under strong photoelectron emission. A possible photoemission example is the sheath on the sunlit side of the moon which exhibits an effective $\gamma$ of about 4 [27]. The scenario cannot be sustained by electron-impact secondary emission [16] or reflections [28] because the emission coefficients only exceed unity at tens-of-eV $T_{et}$ values [29]. Ion-induced emission is negligible in the inverse regime because ions are not accelerated into the target. Field emission [30] cannot occur due to the sign of the electric field at the target when the sheath is inverted.

Cooling the target plasma is of high importance to magnetic fusion energy. It is widely believed that "detachment" will be required to mitigate the divertor plate heat flux problem [31,32]. For detachment, $T_t$ has to be reduced below a few eV, which makes plasma particles recombine and radiate energy. The prevailing method to induce detachment involves injecting impurities that radiate well, but they can contaminate the core [9]. Another path to detachment is to raise collisionality by injection of neutral hydrogen, which helps reduce $T_t$. But injected gas counts against the density limit, compromising stability [31].

The results of this Letter suggest there is an innovative path to detachment by inducing the inverse regime. A target plasma of sub-eV temperature will form as long as the plasma is collisional (typical in SOL's [9]) and the emission is strong enough to maintain an inverse sheath. Recent calculations and sheath simulations by Komm *et al.* suggest that thermionic emission from plasma-heated tungsten in current tokamaks and ITER can reach $\gamma > 1$ below tungsten's melting point of 3695K, despite some redeposition by gyration in the grazing **B** field [1]. But SCL sheaths and a hot target $T_e$ were assumed. Based on the same calculations, we predict that enough emission for an inverse sheath thermionic detachment state can be induced with tungsten, but is unlikely to happen by accident in experiments because a cold detached plasma may not heat the target enough to sustain the requisite thermionic emission. Deliberate heating by an external source might be needed. The required heating and target temperature can be minimized by using low work function cathode materials which emit much more than pure tungsten. A synergistic effect is that once detached, the target plasma density will be reduced due to volume recombination losses, thereby reducing the thermoelectron flux required for an inverse sheath. Further analysis will be needed to understand the properties of inverted detached plasmas and whether implementation in divertors is feasible.

In summary, we analyzed the feedback effect of thermionic emission on the plasma electron temperature in each sheath regime. We showed on theoretical grounds and in kinetic simulations that the transition to the inverse regime cools the target plasma electrons to the sub-eV thermionic temperature. This extreme cooling effect does not take place if the sheath is the SCL type often assumed under strong emission in research areas such as divertors [1], arcs [33], transpiration cooling of hypersonic vehicles [2], dust charging [34], probes [4], and Hall thrusters [25].

This work was performed under the auspices of the U.S. Department of Energy by Lawrence Livermore National Laboratory under Contract No. DE-AC52-07NA27344. G. R. J. was supported in part by the U.S. Department of Energy, Office of Science, Office of Workforce Development for Teachers and Scientists under the Science Undergraduate Laboratory Internship (SULI) Program.